\documentclass[12pt]{article}
\usepackage{a4wide}
\usepackage{amssymb, amsmath,cite}
\usepackage[normalem]{ulem}
\usepackage[utf8]{inputenc}
\usepackage{graphicx}
\usepackage[usenames,dvipsnames]{color}
\usepackage{subfigure}
\usepackage{cancel}
\usepackage[colorlinks]{hyperref}
\usepackage[usenames,dvipsnames]{color}
\hypersetup{
     breaklinks=true,
    pdfstartview={FitH},    
    colorlinks=true,       
    linkcolor=blue,          
    citecolor=red,        
    filecolor=magenta,      
    urlcolor=blue,           
    anchorcolor=green,      
    linktocpage=true
}

\begin{document}
{\renewcommand{\thefootnote}{\fnsymbol{footnote}}
		
\begin{center}
{\LARGE Consistency of Cubic Galileon Cosmology: Model-Independent Bounds from Background Expansion and Perturbative Analyses}

\vspace{1.5em}

Suddhasattwa Brahma$^{1}$\footnote{e-mail address: {\tt suddhasattwa.brahma@gmail.com}} and Md. Wali Hossain$^{2}$\footnote{e-mail address: {\tt mhossain@jmi.ac.in}}\footnote{\tt Present address: Department of Physics, Jamia Millia Islamia, New Delhi-110025, India}
\\
\vspace{0.5em}
$^{1}$ Department of Physics, McGill University, Montr\'eal, QC H3A 2T8, Canada\\
 
$^{2}$ Center for High Energy Physics, Kyungpook National University, Daegu, Korea\\
\vspace{1.5em}
\end{center}
}
	
\setcounter{footnote}{0}

\newcommand{\bea}{\begin{eqnarray}}
\newcommand{\eea}{\end{eqnarray}}
\renewcommand{\d}{{\mathrm{d}}}
\renewcommand{\[}{\left[}
\renewcommand{\]}{\right]}
\renewcommand{\(}{\left(}
\renewcommand{\)}{\right)}
\newcommand{\nn}{\nonumber}
\newcommand{\Mpl}{M_{\textrm{Pl}}}
\newcommand{\lamb}{\Lambda}
\newcommand{\Hinf}{H_{\textrm{inf}}}
\def\H{\mathrm{H}}
\def\V{\mathrm{V}}
\def\e{\mathrm{e}}
\def\be{\begin{equation}}
\def\ee{\end{equation}}
\def\al{\alpha}
\def\bet{\beta}
\def\gam{\gamma}
\def\om{\omega}
\def\Om{\Omega}
\def\sig{\sigma}
\def\lam{\lambda}
\def\ep{\epsilon}
\def\ups{\upsilon}
\def\vep{\varepsilon}
\def\S{\mathcal{S}}
\def\d{\mathrm{d}}
\def\g{\mathrm{g}}
\def\m{\mathrm{m}}
\def\r{\mathrm{r}}

\abstract{We revisit the cosmological dynamics of the cubic Galileon model in light of the recently proposed model-independent analyses of the Pantheon supernova data. At the background level, it is shown to be compatible with data and preferred over standard quintessence models. Furthermore, the model is shown to be consistent with the trans-Planckian censorship conjecture (as well as other Swampland conjectures). It is shown that for the given parametrization, the model fails to satisfy the bounds on the reconstructed growth index derived from the Pantheon data set at the level of linear~perturbations.}

\flushbottom

\section{Introduction}
The $\lamb$CDM flat concordance model, based on General Relativity and the assumption of the Cosmological Principle, has been extremely successful in describing an abundance of cosmological data despite its remarkable simplicity~\cite{Aghanim:2018eyx}. In~this model, the~late-time acceleration of the universe is explained by the presence of a cosmological constant term~\cite{Sahni:1999gb}. Nevertheless, the~true nature of dark energy, responsible for accelerated cosmic expansion, still remains a hotly debated topic~\cite{Riess:2019cxk,Wong:2019kwg,Camarena:2019moy,Riess:2020sih,Macaulay:2013swa,Johnson:2015aaa,Tsujikawa:2015mga,Kazantzidis:2018rnb}. Put differently, the~origin of this exotic non-clustering component, which has a sufficiently negative pressure density to give rise to accelerated expansion, is one of the most interesting outstanding puzzles in cosmology today. Although~the presence of a simple cosmological constant in the Friedmann equations seems to satisfy all available data, there are still several reasons to look for alternative explanations. The~most important challenges to the $\lamb$CDM model comes from theoretical considerations---the very small magnitude of this constant is, in~itself, a~roadblock in understanding this constant as the vacuum energy density~\cite{Martin:2012bt}. Furthermore, recently there have been hints coming in from String Theory that a pure de-Sitter (dS) solution, as~would be the case for the $\lamb$CDM paradigm in the asymptotic future, is not compatible with a ultraviolet (UV) complete theory of gravity~\cite{Vafa:2005ui}. This is a rather remarkable development since it implies that the nature of quantum gravity can leave its imprint on the very late time dynamics of the universe, albeit indirectly through the theoretical consistency of the low energy effective field theory (EFT) being~employed.

Even going beyond String Theory, there have been several quantum gravity arguments which go against the existence of stable dS spacetimes~\cite{dS1,dS2,No_Stringy_dS,Dvali_Exclusion}. In~fact, there have been interesting EFT arguments against having very long periods of accelerated expansion in the early universe dating back several decades~\cite{Martin:2000xs,Brandenberger:2000wr}. The~crux of this argument---the so-called trans-Planckian censorship conjecture (TCC) \cite{Bedroya:2019snp,Bedroya:2019tba}---is that if one has too many $e$-folds of accelerated expansion, one can trace back macroscopic perturbation modes to have their origin in modes which have energies above the Planck scale $\Mpl$, a~fact obviously inconsistent with EFT on curved spacetimes. Although~the origin of this argument lies in analyzing inflationary cosmologies, recent work has found support for this conjecture arising from other well-known facets of String Theory~\cite{Brahma:2019vpl,Cai:2019dzj}. Of~course, the~TCC itself was invoked as a physical reasoning behind the (refined) dS conjecture, namely the idea that there can be no long-lived metastable dS vacua in any EFT which can have a UV-completion of gravity. Leaving aside details for later, all of the weight of this theoretical evidence gives us good reason to look beyond $\lamb$CDM for explaining the nature of dark~energy. 

To construct models beyond the standard $\Lambda$CDM, either we can modify the energy content of the universe~\cite{Wetterich:1987fk,Wetterich:1987fm,Peebles:1987ek,Ratra:1987rm,Copeland:2006wr}, or, we can modify gravity on large cosmological \mbox{scales~\cite{Fierz:1939ix,Horndeski:1974wa,Brans:1961sx,Starobinsky:1980te,Starobinsky:1982ee,Dvali:2000hr,Cartier:2001is,Hwang:2005hb,Nicolis:2008in,deRham:2010kj,DeFelice:2010aj,Clifton:2011jh,deRham:2014zqa,deRham:2012az,Harko:2011kv,Cognola:2007zu,Nojiri:2010wj,Nojiri:2017ncd,Odintsov:2020zkl}}. While the former can be achieved by introducing a slowly rolling scalar field, known as quintessence \cite{Wetterich:1987fk, Wetterich:1987fm,Peebles:1987ek,Ratra:1987rm}, modification of gravity comes in different flavours, e.g.,~Brans--Dicke theory~\cite{Brans:1961sx}, $f(R)$ theories~\cite{Starobinsky:1980te,Starobinsky:1982ee}, Dvali--Gabadadze--Porrati (DGP) model~\cite{Dvali:2000hr}, Galileon models~\cite{Nicolis:2008in}, massive gravity~\cite{Fierz:1939ix,deRham:2010kj}, and so on. The~scalar--tensor theories include a wide range of gravity theories which have a scalar field in addition to the tensor field. All such theories can be represented by the most general scalar--tensor Lagrangian, known as Horndeski theory \cite{Horndeski:1974wa}. Even though the Horndeski Lagrangian contains higher derivation terms, it generates second order equations of motion and thus the theory is ghost-free. Galileon models~\cite{Nicolis:2008in,Deffayet:2009wt, Gannouji:2010au,DeFelice:2010pv, DeFelice:2010nf,Ali:2010gr} are a sub-class of Horndeski theory and preserve Galilean shift symmetry in the flat background where $\phi\to\phi+b_\mu x^\mu+c$, where $b_\mu$ and $c$ are constants. The~recent detection of the event of binary neutron star merger GW170817, using both gravitational waves (GW) \cite{TheLIGOScientific:2017qsa} and its electromagnetic counterpart~\cite{Monitor:2017mdv,GBM:2017lvd} rules out most of the higher derivative terms as they generically predict the speed of GWs to be different from that of the speed of light~\cite{Ezquiaga:2017ekz, Zumalacarregui:2020cjh,Kase:2018aps}. The~only higher derivative term allowed after the GW170817 event is the \mbox{cubic term $\sim G_3(\phi,X)\Box\phi$}, where $G_3$ is a function of the scalar field and its kinetic energy $X\sim (1/2)\partial_\mu\phi\partial^\mu\phi$ (the most general surviving Lagrangian can be found, for~instance, in~\cite{Creminelli:2019kjy}.).~When $G_3\sim X$, the~resulting model is known as cubic Galileon theory~\cite{Chow:2009fm,Silva:2009km,Barreira:2013eea, Bartolo:2013ws,Dinda:2017lpz, Dinda:2018eyt,Zhang:2020qkd} which can have a linear potential. If~we further generalize the potential, the~Lagrangian breaks the shift symmetry but still gives second order equation of motion. These types of models are known as light mass Galileon, or~equivalently as cubic Galileon, models~\cite{Ali:2012cv,Hossain:2012qm, Hossain:2017ica, Brahma:2019kch} which is our scenario under consideration in this paper. Here we should also mention that the kinetic gravity braiding models~\cite{Deffayet:2010qz,Pujolas:2011he} also incorporate the cubic term. Cubic Galileon models without potential do not have a stable late time acceleration~\cite{Gannouji:2010au}. However, in the light mass Galileon models, the late time dynamics is dominated by the scalar field potential and we have a stable late time acceleration. For the same reason we also do not have the instabilities which were discussed in~\cite{Brando:2019xbv}.

Precision cosmological data acts as the most stringent criterion for constraining models of dark energy~\cite{Aghanim:2018eyx}. Recently, a~model-independent analyses of reconstructing the expansion function and the linear growth factor from the Pantheon data set of type-Ia supernovae (SnIa) \cite{Scolnic:2017caz} has been proposed in~\cite{Reconstruction_Pantheon}. Without~referencing any specific model, it was shown how one can derive the expansion function with the assumptions of a spatially homogeneous and isotropic (simply-connected) universe and~that of a smooth expansion rate. With~the additional assumption that Poisson's equation, as~derived from Newtonian gravity, is applicable for the growth of structures on the relevant local scales, one is also able to constrain the linear growth factor from the SnIa data. Combining the value of these two important functions, from~astronomical data alone---namely, the~luminosity-distance measurements in this case---one would thus be able to constrain specific models of dark energy. The~bottom line is that since we are interested in looking beyond $\lamb$CDM as an explanation of dark energy, it is imperative that we use a model-independent analyses to test our theories and the recent work of~\cite{Reconstruction_Pantheon} provides us precisely with~this. 

In this work, we aim to compare the predictions of the cubic Galileon model against the supernova data, both at the level of background evolution and linear perturbations, and~the swampland criterion to set up falsifiability conditions of the model. \mbox{In~Section~\ref{sec:TCC}}, we briefly review the swampland conjectures and, in~particular, the~TCC and their implications for models of late-time cosmology. In~Section~\ref{sec:model}, we go on to introduce our model while choosing parameters so as to satisfy these theoretical constraints. \mbox{In~Section~\ref{sec:data}} we test the predictions of our model for the expansion function and the growth factor, for~the swampland-compatible range of parameter space, against~the model-independent observational constraints. Finally, we conclude in Section~\ref{sec:con} highlighting the advantages of our model against other theories of dark energy such as quintessence and generalized Proca theories~\cite{Heisenberg:2014rta}. 

\section{Swampland Conjectures and Late-Time~Cosmology}
\label{sec:TCC}
It has long been conjectured that not all quantum field theories (QFTs), which are consistent by themselves, cannot be the low-energy limit of some quantum gravity theory~\cite{Palti:2019pca,Brennan:2017rbf}. In~fact, this has been shown to be true for concrete examples~\cite{Kim:2019ths}. For~instance, the~well-known $\mathcal{N}=4$ supersymmetric Yang--Mills theory in $4$-dimensions can be consistently formulated for any arbitrary gauge group G before coupling to gravity. On~the other hand, if~the rank of the G is bigger than $r_{\rm G}>22$, then such theories cannot be consistently coupled to $\mathcal{N}=4$ supergravity in Minkowski space and they belong to the swampland. This example exemplifies what the working definition for the swampland is going to be for us---the set of additional consistency conditions that low-energy EFTs, describing cosmology, must satisfy in order to be consistently completed in the~UV.

Although there are many facets of the swampland, the~most interesting bound for cosmology, in~this context, comes from the so-called dS conjecture~\cite{dS2,Garg:2018reu,Andriot:2018mav}. As~is well-known, dS space plays a key role in our current understanding of both early and late-time description of cosmology. So why is dS space in conflict with string theory? \mbox{Consider some field $\phi$}. Without~gravity, the~range of this field is typically infinite but the swampland distance conjecture says that once the field excursion $|\Delta \phi| > \Mpl$, then there are many (exponentially) light states which descend from the UV, \mbox{destroying an EFT description~\cite{Ooguri:2006in,Baume:2016psm,Klaewer:2016kiy,Blumenhagen:2018nts,Grimm:2018cpv,Grimm:2018ohb,Landete:2018kqf}}. Starting from this observation, it can be shown (say, using entropy arguments~\cite{Bousso:1999xy} or the species bound~\cite{Dvali:2007hz, Dvali:2007wp, Veneziano:2001ah}) that as one goes to these parametrically large regions of field space, the~potential of the field must behave as~\cite{Hebecker:2018vxz}
\begin{eqnarray}
	V \sim e^{-c\phi/\Mpl}\, 
\end{eqnarray}  
for $\phi \gg \Mpl$. Of~course, this is just a generalization of the Dine-Seiberg runaway for moduli fields. Given this, it is easy to derive the (refined) dS conjecture~\cite{dS2}, which goes as
\begin{eqnarray}\label{dSC}
	\dfrac{|V'|}{V} > \dfrac{c}{\Mpl}\,, \hspace{2cm} \text{or}\hspace{2cm} \dfrac{V''}{V} < -\dfrac{\tilde{c}}{\Mpl^2}\,,
\end{eqnarray}
with $c,\tilde{c}$ being $\mathcal{O}(1)$ numbers. These imply that either the potential is steep or it can have an unstable maxima with large tachyonic directions. This, of~course, rules out any meta-stable dS vacua. Furthermore, general arguments from string compactifications also suggest that $\Mpl \,\[|V'|/V\]_\infty \geq \sqrt{2/3}$ \cite{Palti:2019pca}. Of~course, all of these arguments rely on calculation for large distances on field space and one might wonder if they are of any relevance for late-time cosmology. Specifically, one expects field excursions, say of a quintessence field, to~be much smaller than $\Mpl$ and therefore, invalidate the regime where one can use the distance conjecture as the starting point for deriving the (asymptotic) dS~one. 

If the above restrictions on the potential are only applicable in these ``corners'' of moduli space, a~natural question to ask is whether there are any restrictions appearing at all for the potential in the regimes where $|\Delta\phi| \ll \Mpl$. Firstly, note that, keeping in mind the specific behavior the potential must satisfy for large $\phi$, it would be extremely unnatural for the potential to be completely unconstrained far inside moduli space. On~the other hand, the~above restrictions equation~\eqref{dSC} seem to be only strictly applicable in the case of large asymptotic regions. Furthermore, this is where the trans-Planckian censorship conjecture comes in. The~idea here is to restrict the form of the potential, or~equivalently, the~life-span of any dS space, from~general quantum gravity arguments rather than focusing on specific stringy constructions, as~mentioned in the introduction. The~mathematical statement of the TCC is as follows~\cite{Bedroya:2019snp}: no classical macroscopic perturbation with a physical wavelength larger than the Hubble radius should ever originate from trans-Planckian quantum modes, i.e.,
\begin{eqnarray}\label{TCC}
	\dfrac{a_f}{a_i} < \dfrac{\Mpl}{H_f} \;\; \Rightarrow\;\; N < \ln \(\dfrac{\Mpl}{H_f}\)\,,
\end{eqnarray}
where $a(t)$ denotes the scale factor, $H$ the Hubble parameter, and $N$ the number of $e$-foldings, respectively. $i$ and $f$ stand for initial and final times. Without~going into the details of the origin of the TCC (the TCC is really a statement regarding non-unitarity of Hilbert spaces on expanding backgrounds~\cite{Weiss:1985vw,Jacobson:1999zk}. From~this point of view, one can get some $\mathcal{O}(1)$ number refinements of it~\cite{Berera:2020dvn}.), the~remarkable thing is that it produces the dS conjecture equation~\eqref{dSC}, for~asymptotic regions of field space. In~other words, starting from the TCC, one finds

\begin{eqnarray}
	\[\dfrac{|V'|}{V}\]_\infty < \sqrt{\dfrac{2}{3}}\,\dfrac{1}{\Mpl}\,,
\end{eqnarray}
specifying a concrete value for the $\mathcal{O}(1)$ parameter $c$ introduced in equation~\eqref{dSC}. Although~we skip the derivation here, and~refer the interested reader to~\cite{Bedroya:2019snp,Brahma:2019vpl}, this result is true for any potential and one does not have to choose any specific form (such as an exponential one) to arrive at it. More interestingly, starting from the TCC bound on the number of $e$-folds for any accelerating spacetime, it can be shown that metastable dS spacetimes are now allowed for small field excursions with the following upper limit on the lifetime of such dS solutions~\cite{Bedroya:2019snp}
\begin{eqnarray}\label{TCC_1}
	T < \dfrac{1}{\sqrt{V}} \;\ln \(\dfrac{\Mpl}{\sqrt{V}}\)\,.
\end{eqnarray}

Let us now review our main guiding principles coming out of the swampland. Either we give up on dS spacetimes altogether (as suggested by the dS conjecture) or even if we are to have one, it must be extremely short-lived (as suggested by the TCC). The~trouble is that the main approaches to constructing dS spacetimes in string theory, such as the KKLT and the LVS scenarios, all have lifetimes which are exponentially bigger than what the TCC restricts it to be~\cite{Bedroya:2019snp,Westphal:2007xd}, not to mention the numerous obstructions one faces when building an explicit example~\cite{No_Stringy_dS,Dasgupta:2019gcd,Dasgupta:2019vjn}. Furthermore, there are also other indications that meta-stable dS spaces must be short-lived coming from other general arguments~\cite{Dvali:2018jhn,ArkaniHamed:2007ky}. Keeping these in mind, the~lesson for late-time cosmology seems to be to look beyond standard $\Lambda$CDM scenario. However, in~doing so, the~alternative proposals must satisfy the constraints on the potential, as~imposed by the swampland. More concretely, the~specific constraints for us are going to~be:
\begin{itemize}
	\item For the late-time model we are interested in, the~field space excursion is automatically small $\Delta\phi\ll 1$, thereby satisfying the distance conjecture.
	\item For explicit potentials involving scalar fields, equation~\eqref{dSC} must always be satisfied with some number $c$ consistent with the TCC.
	\item Since the lifetime of any consistent dS space is strictly constrained by the TCC equation~\eqref{TCC_1}, it means that our alternate scenario should never asymptote towards a dS attractor.
\end{itemize}

\section{The Cubic Galileon~Model}
\label{sec:model}
Our starting point is to consider the following action with a potential $V(\phi)$ \cite{Ali:2012cv,Hossain:2012qm}
\begin{eqnarray}
\S=\int \d^4x\sqrt{-\g}\Bigl [\frac{\Mpl^2}{2} R-\frac{1}{2}(\nabla \phi)^2\Bigl(1+\frac{\al}{M^3}\Box \phi\Bigr) - V(\phi) \Bigr]+ \S_\m+\S_\r \, ,
\label{eq:action}
\end{eqnarray}
where $M$ is a constant of mass dimension, $\Mpl=1/\sqrt{8\pi G}$ is the reduced Planck mass and  $\al$ is a dimensionless constant. $\S_\m$ and $\S_\r$ correspond to the matter and radiation action, respectively. It is straightforward to see that one can rescale the parameter $\al$ to replace $M$ by $\Mpl$ and reduce the amount of free parameters. This action can be realized as a sub-class of Horndeski theories and one can recover the usual quintessence models on taking $\alpha\rightarrow0$. This type of cubic self-interaction term can also originate from the decoupling limit of the DGP model~\cite{Luty:2003vm,Nicolis:2004qq}.

As mentioned in the introduction, the~primary advantage of the cubic Galileon model is that its action gives rise to late-time acceleration without violating any of the astrophysical constraints especially due to multi-messenger GW astronomy. The~local physics is also preserved through the Vainstein mechanism~\cite{Vainshtein:1972sx}. However, even the cubic Galileon, with~a linear potential and tracker behavior~\cite{Renk:2017rzu} or without potential~\cite{Kimura:2011td}, is strongly disfavored when taking Integrated Sachs-Wolfe effect (ISW) measurements into account~\cite{Renk:2017rzu}. On~the other hand, for~thawing dynamics, when the Galileon scalar field dynamics is mostly dependent on the potential, the~constraint coming from ISW can be relaxed~\cite{Kase:2018aps}. For~our case, we shall introduce a potential other than the linear one, by~introducing an exponential potential. Firstly, note one needs to add such a potential term in order to get ghost-free, stable late-time acceleration in cubic Galileon~\cite{Ali:2010gr,Gannouji:2010au}. Moreover, if~our goal is to show that the inclusion of higher derivative terms shall lead to dark energy models which can be made to obey both observational constraints and the swampland conjectures, this is a dual task that is not fulfilled by simpler quintessence models~\cite{Heisenberg:2018yae,Heisenberg:2020ywd,Akrami:2018ylq}. We shall elaborate on this later. Furthermore, quantum corrections typically give rise to the appearance of higher derivative terms in Galileon theories when the non-renormalizable theorem is violated~\cite{Pirtskhalava:2015nla}, when the Galileon symmetry is not weakly-broken. Finally, the~form of the potential is chosen to be an exponential one since it is the \textit{least-constrained} form of scalar potential, as~per the swampland~\cite{Agrawal:2018own}.

The background Einstein equations for the cubic Galileon model can be obtained by varying the action equation~(\ref{eq:action}) with respect to (w.r.t.) the metric $\g_{\mu\nu}$:  

\begin{eqnarray}
3M_{\rm{pl}}^2H^2 &=&\rho_\m+\rho_\r + \rho_\phi\,,\label{Friedmann1}\\
M_{\rm{pl}}^2(2\dot H + 3H^2)&=&-\frac{\rho_\r}{3} - P_\phi\,,\label{Friedmann2}
\end{eqnarray}
where
\begin{eqnarray}
\rho_\phi &=& \frac{\dot{\phi}^2}{2}\Bigl(1-\frac{6\alpha}{M^3} H\dot{\phi}\Bigr) + V{(\phi)} \, ,
\label{eq:rhopi}\\
P_\phi &=& \frac{\dot{\phi}^2}{2} \Bigl(1+\frac{2\alpha}{M^3}\ddot{\phi}\Bigr) - V(\phi) \, .
\label{eq:ppi}
\end{eqnarray}
where $\rho_\m$ and $\rho_\r$ are the energy densities of non-relativistic matter ($P_\m = 0$) and radiation ($P_\r = \rho_\r/3$) respectively. The~background EOM for the Galileon field is similarly given by
\begin{eqnarray}
\ddot{\phi}+ 3H\dot{\phi}- \frac{3\alpha}{M^3} \dot{\phi}\Bigl(3H^2\dot{\phi} + \dot{H}\dot{\phi} + 2H\ddot{\phi}\Bigr)+ V'(\phi)=0 \, ,
\label{eq:phi_eom}
\end{eqnarray}
where we have everywhere assumed a spatially flat, Friedmann--Lema\^itre--Robertson--Walker (FLRW) metric $\d s^2 = -\d t^2 + a^2(t)\d r^2$, the~spatially flat assumption being the same as what has been assumed for the model-independent reconstruction of the expansion function and the growth index from supernovae~data.

The background cosmological dynamics is governed by the following autonomous system of equations~\cite{Ali:2012cv,Hossain:2012qm}:
\begin{align}
\frac{{\rm d}x}{{\rm d}N}&=x\Bigl(\frac{\ddot{\phi}}{H\dot{\phi}}-\frac{\dot H}{H^2}\Bigr)
\label{eq:x}\\
\frac{{\rm d}y}{{\rm d}N}&=-y \Bigl(\sqrt{\frac{3}{2}}\lambda x+\frac{\dot H}{H^2}\Bigr)
\label{eq:y}\\
\frac{{\rm d}\epsilon}{{\rm d}N}&=\epsilon \Bigl(\frac{\ddot{\phi}}{H\dot{\phi}}+\frac{\dot H}{H^2}\Bigr)
\label{eq:ep}\\
\frac{{\rm d}\Omega_r}{{\rm d}N}&=-2\Omega_r\Bigl(2+\frac{\dot H}{H^2}\Bigr)
\label{eq:omr}\\
\frac{{\rm d}\lambda}{{\rm d}N}&=\sqrt{6}x\lambda^2(1-\Gamma)
\end{align}
where $N=\text{ln}\, a$ is the number of $e$-folds. In~order to bring the evolution equations to this form, we have defined the dimensionless quantities
\begin{eqnarray}
x &=& \frac{\dot{\phi}}{\sqrt{6}H M_{\rm{pl}}}\,, 
\label{eq:x1}\\
y &=& \frac{\sqrt{V}}{\sqrt{3} H M_{\rm{pl}}}\,,
\label{eq:y1}\\
\epsilon &=& -6\frac{\alpha}{M^3}H\dot \phi\,,
\label{eq:ep1}\\
\Omega_\r &=& \frac{\rho_\r}{3 \Mpl^2 H^2}\,,
\label{eq:omr1}\\
\lambda &=& -M_{\rm{pl}}\frac{V'}{V},
\label{eq:lam1}
\end{eqnarray}
and the equation-of-state (EoS) parameters,
\begin{eqnarray}
w_{\rm eff} &=& -1-\frac{2}{3}\frac{\dot H}{H^2} = \frac{3 x^2 \(4+8\ep+\ep^2\)-2\sqrt{6}xy^2\ep\lam-4(1+\ep)\(3y^2-\Om_\r\)}{3\(4+4\ep+x^2\ep^2\)} \, , \\
w_\phi &=& \frac{P_\phi}{\rho_\phi} = -\frac{12 y^2 (1+\epsilon )+2 \sqrt{6} x y^2 \ep\lam-x^2 \(12+24\ep+\ep^2(3-\Om_\r)\)}{3\left(4+4 \epsilon +x^2 \epsilon ^2\right) \left(y^2+x^2 (1+\epsilon )\right)} \, ,
\end{eqnarray}
where $w_{\rm eff}$ and $w_\phi$ are the effective and scalar field EoS. The~parameter $\Gamma = VV_{,\phi\phi}/V_{,\phi}^2$ which, for~our choice of an exponential potential,
\begin{equation}
 V(\phi) = V_0 e^{-\lambda \phi/\Mpl} \, ,
 \label{eq:pot}
\end{equation}
implies $\Gamma=1$ since $\lambda= \text{const.}$. The~background Friedmann Equations \eqref{Friedmann1} and \eqref{Friedmann2}, can be used to obtain
\begin{align}
\frac{\dot H}{H^2}&=\frac{2(1+\epsilon)(-3+3y^2-\Omega_r)-3x^2(2+4\epsilon+\epsilon^2)+\sqrt{6}x\epsilon y^2\lambda}{4+4\epsilon+x^2\epsilon^2} \,,\\
\frac{\ddot{\phi}}{H\dot{\phi}}&=\frac{3x^3\epsilon-x\Bigl(12+\epsilon (3+3y^2-\Omega_r)\Bigr)+2\sqrt{6}y^2\lambda}{x(4+4\epsilon+x^2\epsilon^2)} \,.
\end{align}

Rewriting the matter density $\Omega_\m := \rho_\m/(3\Mpl^2 H^2)$, one can express  the constraint equation as $\Omega_\m+\Omega_\r+\Omega_\phi=1$ where
\begin{equation}
\Omega_\phi=x^2(1+\epsilon)+y^2 \,,
\label{eq:ompi}
\end{equation}
where we have introduced the density parameter of the Galileon field as $\Omega_\phi :=\rho_\phi/(3\Mpl^2H^2)$. 

\begin{figure}
\centering
\includegraphics[scale=.65]{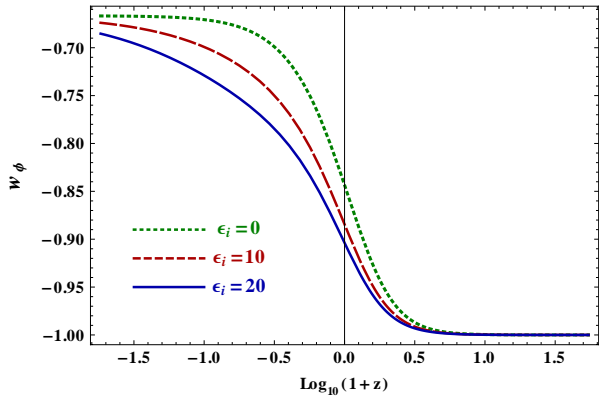}
\caption{Evolution of scalar field EoS has been shown. Green (dotted), red (dashed) and blue (solid) curves correspond to $\ep_i=0,\; 10,\; 20$ respectively for $\lam=1$ and $\Om_{\m0}=0.3$.} 
\label{fig:wpi1}
\end{figure}

The fixed point analyses for the cubic Galileon model, with~an exponential potential, has been extensively studied elsewhere~\cite{Brahma:2019kch} and we only quote a few results here. Firstly, note that the parameter $\epsilon$ denotes the deviation of this system from standard quintessence scenario, as~evidenced from Equations \eqref{eq:rhopi}--\eqref{eq:phi_eom} and \eqref{eq:ep1}. From~Equation~\eqref{eq:phi_eom} it can be understood that the contribution from the higher derivative term in the cubic Galileon Lagrangian can act as a frictional term. This can be achieved by considering positive initial values of $\ep$. In~Figure~\ref{fig:wpi1} we have shown this effect for different initial values of $\ep$ i.e.,~$\ep_i$ and we can see that as we increase the value of $\ep_i$ the scalar field EoS, at~present, moves towards the de Sitter case. This freedom makes this model more consistent with both observation and string swampland conjectures than the quintessence model~\cite{Hossain:2012qm,Brahma:2019kch}. Similar behavior can be obtained by changing the initial value of $x$ for non-zero $\ep_i$ \cite{Brahma:2019kch}.

\begin{figure}
\centering
\includegraphics[scale=.65]{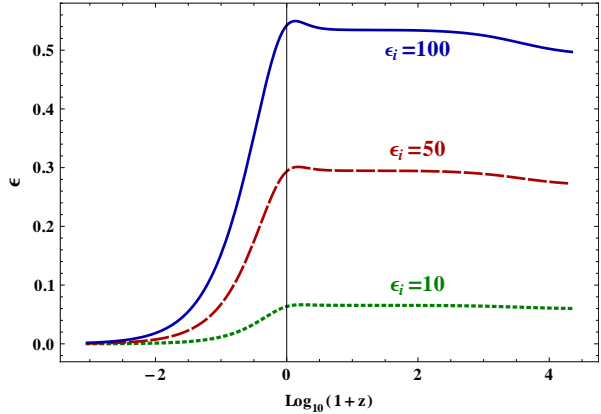}
\caption{Evolution of $\ep$ has been shown. Dotted, dashed, and solid curves correspond to $\ep_i=10,\; 50,\; 100$ respectively.} 
\label{fig:epsilon}
\end{figure}

Now, although~the overall number and character of the fixed points of this model differ with those for quintessence~\cite{Copeland:1997et}, at~late times, this model has an attractor point exactly the same as the one for quintessence, labeled by the following values: $\left(\Omega_\phi = 1,\, \Omega_m=0,\, \ep= 0, \right. \\ \left. \,x= \lambda/\sqrt{6},\, y = \sqrt{1-\lambda^2/6}\right)$. Interestingly, the~most important property of this fixed point of the Galileon model is that $\ep$ flows to zero on the four-dimensional phase space which we have shown in Figure~\ref{fig:epsilon} for different values of $\ep_i$. Figure~\ref{fig:wpi1} also shows that the cubic Galileon model reduces to the the quintessence model in the~future.

At leading order, the~EoS, at~this attractor point, is given by $\left(w_\phi=-1+\lambda^2/3 = w_{\rm eff}\right)$. The~crucial feature of this attractor fixed point, from~the point of view of the TCC, is that it is distinct from the dS attractor solution. The~dS critical point is characterized by $(\Omega_\phi = 1,\, \Omega_m=0,\, \ep= 0, \,x= 0,\, y=1)$ and the late-time fixed point of our model approaches the dS one only in the limit $\lambda\rightarrow 0$. Therefore, since the TCC rules out a dS attractor solution, future evolution in our model is completely safe and navigates clear of the swampland, especially in light of the $\mathcal{O}(1)$ number $\lambda$ which we shall require of our solution later~on.

\section{Comparison with Model-Independent Analyses of Snia~Data}
\label{sec:data}
Having introduced our model, we now need to specify the observable constraints that we shall require it to satisfy in order to be phenomenologically viable. Naturally, this is being  done after we imposed the theoretical swampland constraints in the previous section to make our model already consistent with quantum gravity. To~compare the scenario under consideration with the Pantheon sample of SNIa observation~\cite{Scolnic:2017caz}, we will follow the approach considered in~\cite{Reconstruction_Pantheon,Heisenberg:2020ywd}. In~\cite{Heisenberg:2020ywd}, the~authors reconstructed two functions, $q(a)$ and $\gam(a)$, corresponding to the background expansion and the growth index of linear perturbations, respectively.

For a homogeneous and isotropic background, one can define the cosmic expansion function $E(a)$ in terms of the Hubble parameter $H(a)$, $a$ being the scale factor of the FLRW metric, as
\begin{eqnarray}\label{Expansion_Function}
E^2(a) := \(\Omega_{\r 0} a^{-4} + \Omega_{\m 0} a^{-3}+ \Omega_{\rm DE}(a)\) = \dfrac{H^2(a)}{H^2_0}\,,
\end{eqnarray} 
where $H_0$ is the Hubble parameter today. The~$\Omega$'s denote the energy density corresponding to radiation, matter, and dark energy and subscripts $0$ represent energy densities at their present values. For~$\Lambda$CDM, one would replace the time-dependent dark-energy density parameter $\Omega_{\rm DE}(a)$ by the value of the cosmological constant density $\Omega_{\Lambda 0}$ today. However, the~above expression is more general and allows us to calculate the expansion function given a general dark energy~model.

Once we assume a homogeneous and isotropic universe, then the background dynamics must be described by a single degree of freedom. Without~resorting to any specific functional form for the cosmic expansion factor, which would require some specific underlying theoretical model, the~authors of~\cite{Reconstruction_Pantheon} reconstruct $E(a)$ using the luminosity-distance measurements in Pantheon SN-sample of type Ia supernovae~\cite{Scolnic:2017caz}. We do not go into the details of this analyses here, which has as its input the redshifts $z$ and distance moduli $\mu$ of individual supernovae and then uses shifted Chebyshev polynomials for the expansion. We refer the reader to the original work~\cite{Reconstruction_Pantheon} for details. For~our purposes, we want to use the findings of the model-independent reconstruction of the expansion function to see if our late-time cosmology model is able to fit the data. We emphasize that we do not wish to carry out the model-independent reconstruction or~review the analyses but~rather just to check whether the predictions of our model fall within the constraints derived from such a procedure. More specifically, since we are interested in the late-time dynamics, we define a function $q(a)$, following~\cite{Heisenberg:2020ywd}
\begin{equation}
q(a):=\dfrac{E^2(a)-\Om_{\m0} a^{-3}}{\(1-\Omega_{\m 0}\)}\,,
\end{equation}
such that we can plot this function from the reconstructed $E(a)$. For~the standard $\Lambda$CDM universe, $q(a)=1$. For~our cubic Galileon model, the~dark energy density $\Om_{\rm DE}=\Om_\phi$.

\begin{figure}
	\centering
	\includegraphics[scale=.77]{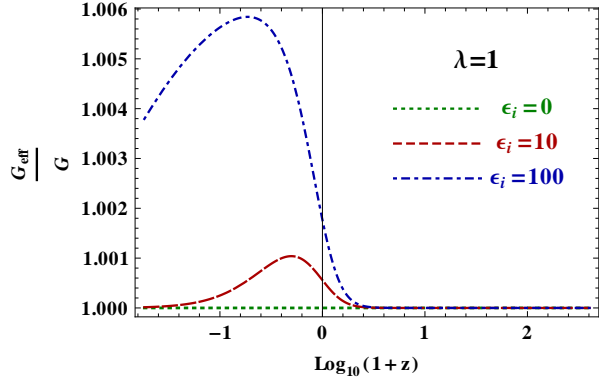}
	\caption{Evolution of $G_{\rm eff}$ has been shown. Green (dotted), red (dashed), and blue (dot-dashed) curves correspond to $\ep_i=0,\; 10,\; 100$ respectively for the exponential potential with slope $\lambda=1$.} 
	\label{fig:Geff}
\end{figure}

Just as the expansion function can be a powerful observable to constrain a dark energy model, similarly one can also use the growth function of linear density perturbations to do the same. To~define the growth index $\gam(a)$ let us first consider the following perturbed metric in the Newtonian gauge
\begin{equation}
\d s^{2} = -(1+2\Psi) \d t^{2} + a(t)^{2}(1-2\Phi) \d r^2,
\label{eq:pert_metric}
\end{equation}
where $\Psi$ and  $\Phi$ are the scalar potentials. For~the cubic Galileon, $\Psi=\Phi$ \cite{Ali:2012cv,Hossain:2017ica}, i.e.,~there is no anisotropic stress. Now, as~we are interested in the scales which are large but sub-Hubble we will impose the usual subhorizon ($k^2\gg H^2$) and quasistatic ($|\ddot\phi|\lesssim H|\dot\phi|\ll k^2|\phi|$) approximations. Under~these assumptions, the~evolution equation of the matter density contrast $\delta_\m=\delta\rho_\m/\rho_\m$, where $\delta\rho_\m$ is the perturbation in matter energy density, in~Fourier space, for~the action equation~(\ref{eq:action}), becomes~\cite{Ali:2012cv,Hossain:2017ica}
\begin{eqnarray}
\ddot\delta_\m+2H\dot\delta_\m-\frac{3}{2} \frac{G_{\rm eff}}{G}\Om_\m H^2\delta_\m=0 \, ,
\label{eq:den_con_evo}
\end{eqnarray}
where the effective Newton's constant
\begin{eqnarray}
G_{\rm eff} &=& G\frac{2P}{2P-Q^2} \, , \\
P &=& 1-\frac{2\al}{M^3}\(\ddot\phi+2H\dot\phi\) \, , \\
Q &=& \frac{\al}{M^3\Mpl}\dot\phi^2 \, .
\end{eqnarray}
where $G$ is the Newton's constant. For~$\Lambda$CDM as well as for the minimally coupled quintessence field, $G_{\rm eff}=G$. Therefore, the~structure of Equation~(\ref{eq:den_con_evo}) looks the same for both $\Lambda$CDM and quintessence but the difference is incorporated in the expansion history $H(a)$. For~cubic Galileon, the~difference is also incorporated in the evolution of the effective Newton's constant ($G_{\rm eff}$). In~Figure~\ref{fig:Geff}, we have shown the evolution of $G_{\rm eff}$ in cubic Galileon model and we can see that, at~present, the~model can behave rather differently from quintessence even at the level of linear perturbations. In the future, the~model merges with the quintessence model and $G_{\rm eff}$ becomes the Newton's constant $G$. 

Since $\delta_m(t,\vec k)$, commonly referred to as the density contrast, satisfies a homogeneous Equation~(\ref{eq:den_con_evo}), it can be separated into a $t$-dependent and a $k$-dependent part as $\delta_m(x,t) = D(t) \delta(0,\vec k)$, where $\delta(0,\vec k)$ is related to the primordial density fluctuations and $D(t)$ satisfies the following evolution equation
\begin{eqnarray}\label{GG_2}
D'' + \(2 + \dfrac{H'}{H}\) D' - \dfrac{3}{2}\frac{G_{\rm eff}}{G}\Omega_{\m} D = 0\,.
\end{eqnarray}
where prime denotes a derivative with respect to $\ln a$. Out of the two linearly-independent solutions of this equation, we focus on the growing solution denoted by $D_+(a)$ from which we can define the growth factor as
\begin{equation}
f(a)=\frac{\d \ln D_+}{\d \ln a} \,,
\end{equation}
which satisfies the following evolution equation:
\begin{eqnarray}
f'+f^2+\frac{1}{2}(1-3w_{\rm eff})f-\frac{3}{2}\frac{G_{\rm eff}}{G}\Om_\m=0 \, .
\end{eqnarray}

Finally, the~growth index $\gam$ is related to the growth factor $f$ through the following parametrization~\cite{Peebles:1984ge,Lahav:1991wc}
\begin{eqnarray}\label{GI_1}
f(a) = \Omega_\m^{\gamma(a)}(a)\,,
\end{eqnarray}

$\gam$ generally varies very slowly with redshift $z$ (though there are models where it can vary significantly, e.g.,~$f(R)$ models~\cite{Belloso_2011}). For~$\Lambda$CDM, $\gam=0.554$ at $z=0$ and $\gam=0.545$ for large $z$ and $\Om_{\m0}=0.3$. Therefore, sometimes we approximate $\gam=6/11$ as a constant for $\Lambda$CDM~\cite{Peebles:1984ge,Lahav:1991wc}. Considering this slow variation in $\gam$ it has been parameterized earlier in terms of the dark energy EoS~\cite{Wang:1998gt}. In~Reference~\cite{Reconstruction_Pantheon}, $\gam$ has been approximately parametrized in terms of the Hubble parameter and the matter density parameter as
\begin{eqnarray}
\gamma = \dfrac{\beta + 3\omega}{2\beta + 5\omega}\,,
\label{eq:gam}
\end{eqnarray}
where
\begin{eqnarray}
\beta &=& 3 + 2 \,\dfrac{\d \ln E}{\d \ln a}=-\frac{\d\ln\Om_\m}{\d\ln a} \,, \\ 
\omega &=& 1- \Omega_{\m}\,, 
\label{eq:omega1}\\
\Om_\m &=& \frac{\Om_{\m0} a^{-3}}{E^2} \, .
\end{eqnarray}

\begin{figure}
\centering
\subfigure[\textbf{Green} (dotted), \textbf{red} (dashed), and \textbf{blue} (dot-dashed) curves correspond to \mbox{$\ep_i=0,\; 10,\; 100$} respectively for the exponential potential with slope $\lambda=0.8$.]{\includegraphics[scale=.7]{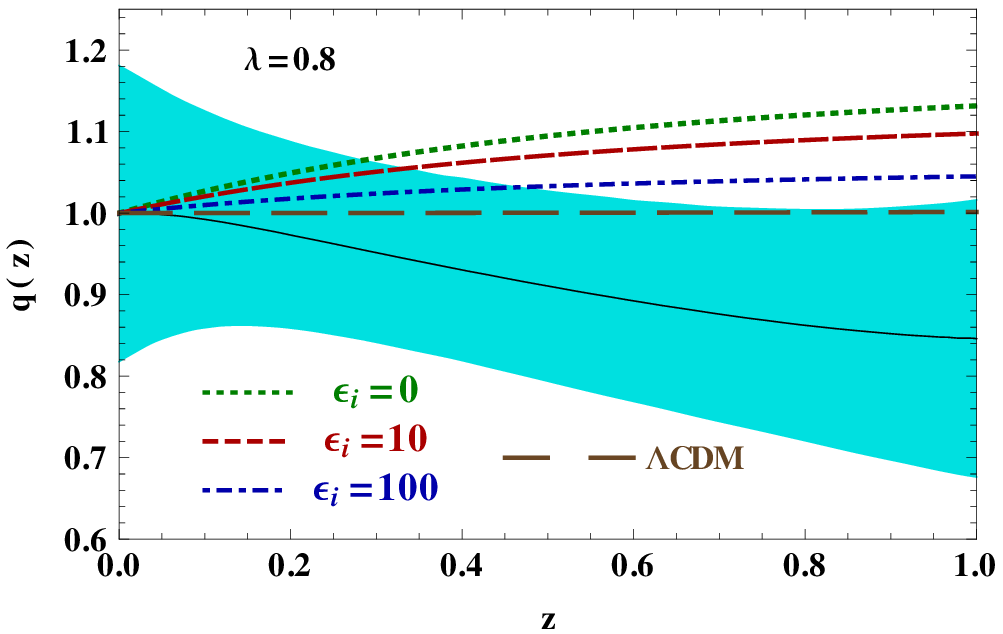}\label{fig:q_ep}}
~~~~~~~~~~~~~~
\subfigure[\textbf{Green} (dotted), \textbf{red} (dashed), and \textbf{blue} (dot dashed) curves correspond to \mbox{$\lam=0.4,\; 0.7,\; {\rm and}\; 1$} respectively. We have kept $\ep_i=10$.]{\includegraphics[scale=.7]{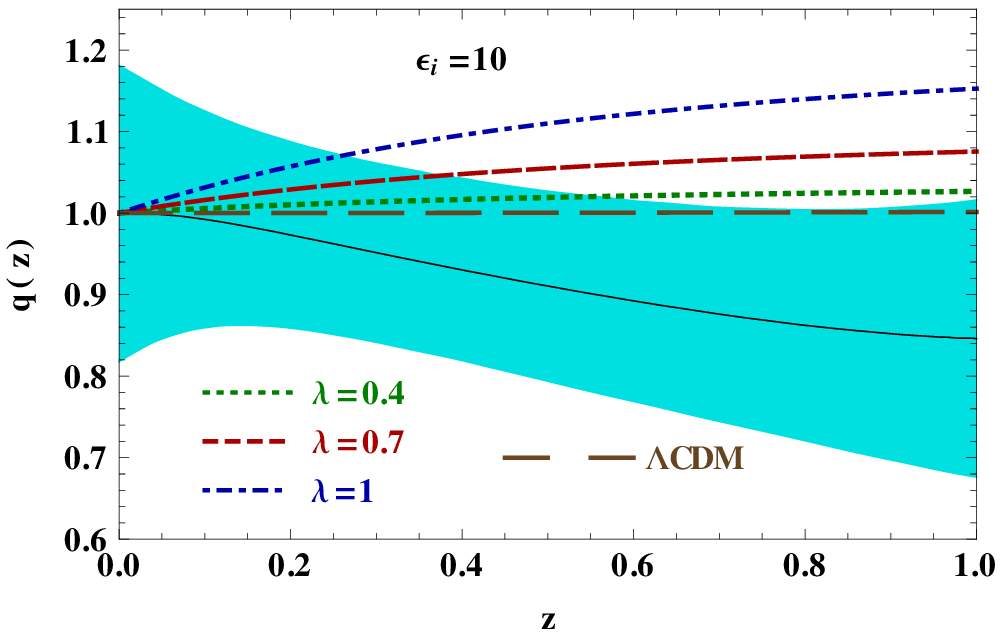}\label{fig:q_lam}}
\caption{Evolution of $q(z)$ has been shown. The~shaded region is the $1\sig$ uncertainty in the reconstructed $q(z)$ from the Pantheon data set of SNIa~\cite{Scolnic:2017caz}. The~brown (long-dashed) line represents $q(z)$ for $\Lambda$CDM. For~both the figures $\Om_{\m0}=0.3$.} 
\label{fig:q}
\end{figure}

The interesting thing to note is that once the cosmic expansion function $E$ is reconstructed from supernovae data in a model-independent way, one can use that to reconstruct the growth index $\gam$, using Equation~(\ref{eq:gam}), and~put observational constraints on it in a model-independent manner as well. Furthermore, this is what has precisely been done in Reference~\cite{Reconstruction_Pantheon,Heisenberg:2020ywd} using the Pantheon data set of SNIa~\cite{Scolnic:2017caz}. In~this paper, we use the reconstructed $E$ and $\gam$ from the Reference~\cite{Reconstruction_Pantheon,Heisenberg:2020ywd} and compare our models with the Pantheon data set of~SNIa.

In Figure~\ref{fig:q}, we have shown the evolution of $q(z)$ for $\lam=0.8$. The~shaded region in the plot is the $1\sig$ uncertainties of the reconstructed $q(z)$ from the Pantheon data set of SNIa. From~the figure, it is clear that the cubic Galileon model can be more suitable than the quintessence model as we increase the value of $\ep_i$. At~the level of background evolution, the~cubic Galileon model can accommodate a larger value for $\lam$, as~required by the swampland, and~yet be completely consistent with data. Importantly, the~future fixed point for the Galileon model is \textit{not} a dS attractor, as~this is the case for some other dark energy models such as Proca theories~\cite{DeFelice:2016yws}. This makes this model consistent with the TCC, as~described~earlier.

Let us now compare the growth index  $\gam(z)$ from this model with data. Before~doing so, for~low redshift, we need to rewrite the Equation~\eqref{eq:gam} in terms of the scalar field~EoS~by~using
\begin{equation}
 w_\phi=\frac{1}{3(1-\Om_\m)}\frac{\d\ln\Om_\m}{\d\ln a} \, ,
\end{equation}
and Equation~\eqref{eq:omega1} as
\begin{equation}
 \gam=\frac{3(w_\phi-1)}{6w_\phi-5} \, .
 \label{eq:gam1}
\end{equation}

\begin{figure}
\centering
\subfigure[ \textbf{Green} (dotted), \textbf{red} (dashed), and \textbf{blue} (dot-dashed) curves correspond to \mbox{$\ep_i=0,\; 10,\; 100$} respectively for the exponential potential with slope $\lambda=0.8$.]{\includegraphics[scale=.7]{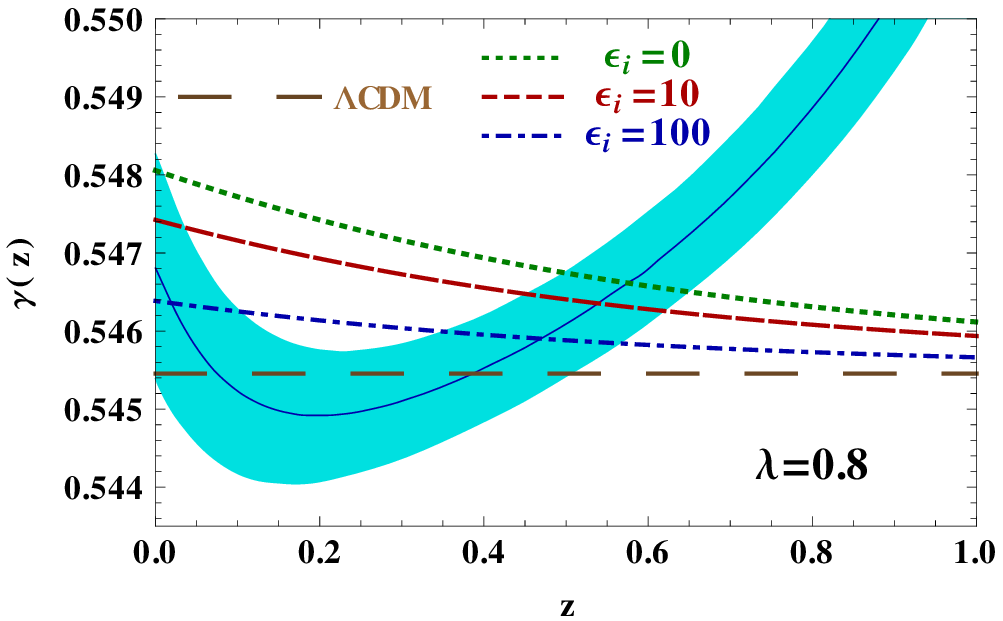}\label{fig:gam_ep}}
~~~~~~~~~~~~~~
\subfigure[ \textbf{Green} (dotted), \textbf{red} (dashed), and \textbf{blue} (dot dashed) curves correspond to \mbox{$\lam=0.4,\; 0.7,\; {\rm and}\; 1$} respectively. We have kept $\ep_i=10$.]{\includegraphics[scale=.7]{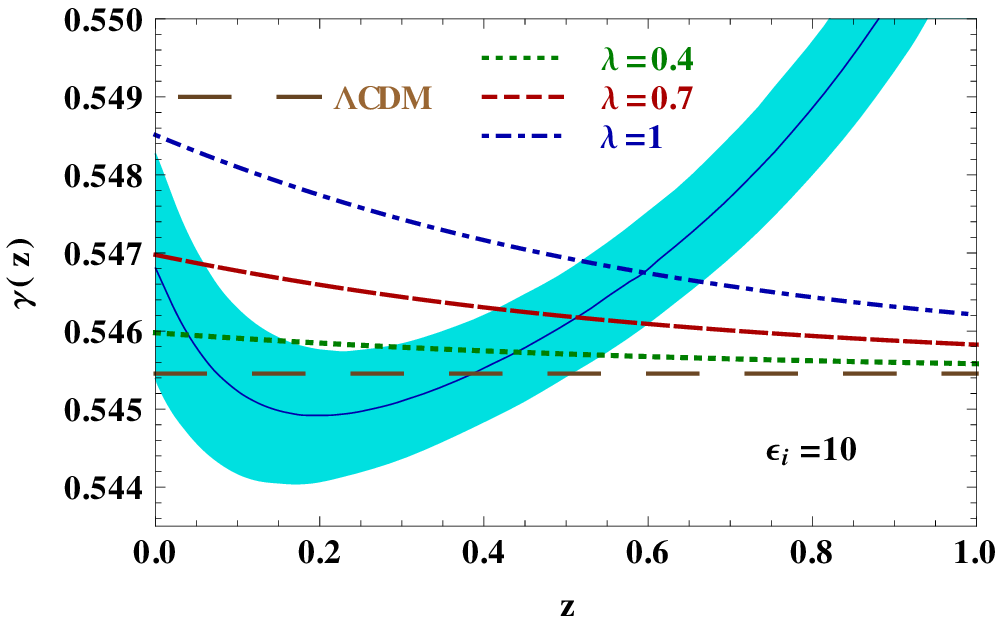}\label{fig:gam_lam}}
\caption{Evolution of $\gam(z)$ has been shown. The~shaded region is the $1\sig$ uncertainty in the reconstructed $\gam(z)$ from the Pantheon data set of SNIa~\cite{Scolnic:2017caz}. The~brown (long-dashed) line represents $\gam(z)$ for $\Lambda$CDM. For~both of the figures $\Om_{\m0}=0.3$.} 
\label{fig:gam}
\end{figure}

Equation~\eqref{eq:gam1} was first calculated in Reference~\cite{Lahav:1991wc} for a constant EoS $w$ and for $\Lambda$CDM we get $\gam=6/11$. Here we should also note that both the Equations~\eqref{eq:gam} and \eqref{eq:gam1} are approximations for the growth index $\gam(z)$.

As is clear from Figure~\ref{fig:gam}, the~$\gam(z)$ for the cubic Galileon model fails to satisfy the $1-\sig$ bound of the reconstructed data, for~all $z$. Although~the predictions of our model fit the data better at the present time than quintessence, it does not have the right shape to be consistent at large redshifts. Firstly, note that this constitutes a falsifiability condition for the cubic Galileon model, emerging at the level of perturbations. This is a new result for the model, in~contrast to the features that have been studied thus far~\cite{Brahma:2019kch}. Nevertheless, one should also keep in mind that Equation~\eqref{eq:gam}, which has been extensively used in the model-independent reconstruction, was derived using the approximation $\ln\Omega_m = \ln \left(1-\omega\right) \approx -\omega$ \cite{Reconstruction_Pantheon}, which should probably be valid only at low $z$. Therefore, one should reconstruct the growth index without making this assumption to see if the results hold at large redshifts. Finally, note that if the model-independent reconstruction of the growth index \textit{does} hold true at large $z$, then even the $\Lambda$CDM model is disfavoured according to~this. 

\section{Conclusions}
\label{sec:con}
In this work, we present some theoretical motivations to look beyond the standard $\Lambda$CDM paradigm and look at alternative models of dark energy. Specifically we consider the cubic Galileon model, which has not yet been ruled out by multimessenger gravitational wave observations~\cite{Ezquiaga:2017ekz,Zumalacarregui:2020cjh,Kase:2018aps}. We find that the cubic Galileon model fits the background data much better than quintessence while at the same time being consistent with the swampland conjectures. However, at~the level of linear perturbations, the~model fails to satisfy the bounds on the reconstructed growth index derived from Pantheon data set of SNIa~\cite{Scolnic:2017caz}.

One way to satisfy the bounds on the growth index might be to look at models of dark energy allowing for a phantom regime. In~\cite{Heisenberg:2020ywd}, Proca theories were considered, which is an example of such models. However, although~such theories do fit the data much better, the~future fixed point of it is an asympototically dS spacetime, which is ruled out by the TCC due to theoretical consistency. Therefore, our future goal would be to look for models which, even on allowing for phantom regimes, do not have dS fixed points like in the case of quintessence. These might be able to simultaneously fit the data as well as be consistent with the~swampland.

Given the myriad of alternative theories of dark energy, it is important to make progress by checking these theories against both theoretical and observational consistency conditions. In~this context, it is interesting to note how linear perturbation theory can test the viability of a given cosmological model in different ways. In~this paper, we impose the TCC---a condition which arises from the non-unitarity of the Hilbert space of linear perturbations on an accelerating background---as a theoretical consistency for a dark energy model. On~the other hand, the~growth index, parametrizing the growth of linear perturbations, turns out to be a crucial feature in constraining dark energy models from observed data. We believe that, moving forward, this reconstructions of background expansion functions, together with that for the growth index, would be extremely useful in constraining other (theoretically) well-motivated models of dark~energy.
\section*{Acknowledgements}
We are grateful to Matthias Bartelmann for helpful correspondence regarding the reconstruction procedure. SB also thanks Robert Brandenberger for stimulating discussions and for helpful comments on an earlier version of this draft. SB is supported in part by the NSERC (funding reference CITA 490888-16) through a CITA National Fellowship and by a McGill Space Institute fellowship. MWH is supported by the National Research Foundation of Korea (NRF No-2018R1A6A1A06024970) funded by the Ministry of Education, Korea.

\bibliographystyle{hunsrt}
\bibliography{references}
\end{document}